# Evaluation of Giga-bit Ethernet Instrumentation for SalSA Electronics Readout (GEISER)

Gary S. Varner*, Laine Murakami, David Ridley, Chaopin Zhu and Peter Gorham

*Contact: varner@phys.hawaii.edu
**Instrumentation Development Laboratory**
Department of Physics and Astronomy, University of Hawaii at Manoa

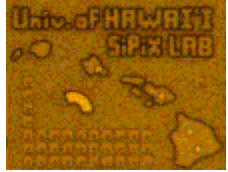

**ABSTRACT**

An instrumentation prototype for acquiring high-speed transient data from an array of high bandwidth antennas is presented. Multi-kilometer cable runs complicate acquisition of such large bandwidth radio signals from an extensive antenna array. Solutions using analog fiber optic links are being explored though are very expensive. We propose an inexpensive solution that allows for individual operation of each antenna element, operating at potentially high local self-trigger rates. Digitized data packets are transmitted to the surface via commercially available Giga-bit Ethernet hardware. Events are then reconstructed on a computer farm by sorting the received packets using standard networking gear, eliminating the need for custom, very high speed trigger hardware. Such a system is completely scalable and leverages the billions of development dollars already invested by the telecommunications industry. Test results from a demonstration prototype are presented.

## 1. Motivation

Detection of Ultra High Energy (UHE) neutrino interactions requires immense effective target volumes due to the miniscule flux of such highly energetic particles. Recently a number of initiatives [1,2] have been proposed to exploit the electromagnetic impulsive signature for UHE neutrino interactions, named in honor of Askaryan [3] who predicted the phenomenon. Any dense material may serve as a target, however for the shower induced radio impulse to be observable, a radio transparent target material is required. Two candidates that have been extensively explored are ice and salt[4]. A concept drawing for an antenna array is shown in Figure 1. The readout system presented here can work with either media though the extreme low temperature operating environment of the Antarctic ice [1] makes the off-the-shelf component solution less attractive. Therefore the initial effort has focused on developing a prototype for doing an exploratory borehole test in a salt dome. In order to detect the pulses of interest with high sensitivity, direct digitization of a large bandwidth is required. Also, the ability to trigger with thresholds right at the thermal noise limit is required. A full-custom CMOS integrated circuit has been designed to meet these requirements[5]. The architecture of this Self-Triggered Recorder for Analog Waveforms (STRAW) contains functionality for both triggering as well as sampling the RF pulses, as indicated in Figure 2.

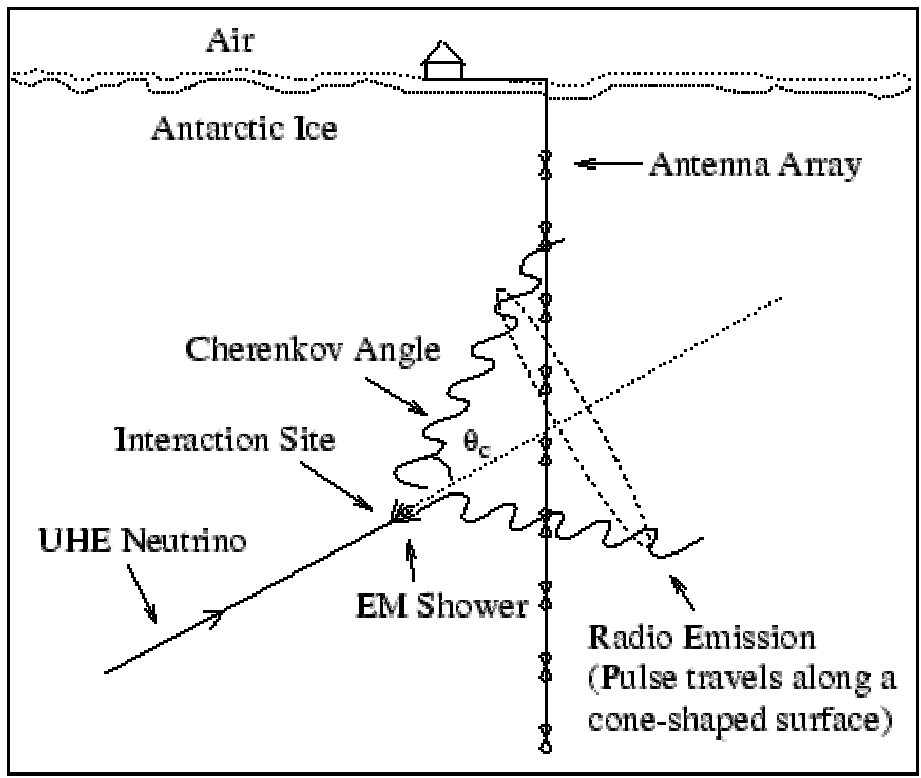

**Figure 1:** Concept drawing of the Radio ICE (RICE) configuration. A Salt Sampling Array (SalSA) is conceptually identical though represents a larger effective target volume due to the higher density of rock salt compared to ice.

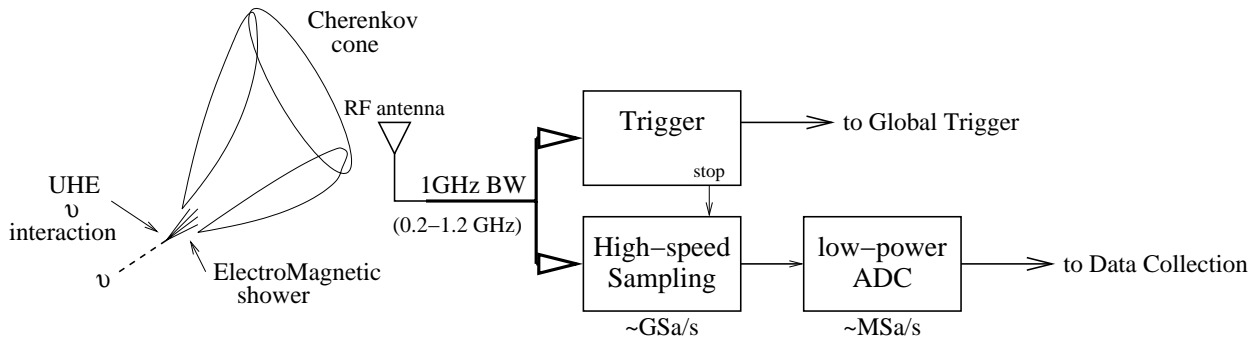

**Figure 2:** Overview of the Self-Triggered Recorder for Analog Waveforms (STRAW) architecture, indicating the stand-alone capability to observe UHE neutrino interactions.

A number of readout architectures have been considered, including direct copper cable transmission, analog fiber optic link and high-speed differential copper links, both digital and analog. Since the transmission distances can be many kilometers, frequency dependent attenuation can become severe. To minimize this, higher quality cable may be used, though the bundle thickness becomes an issue, especially since it must pass through all subsequent antennae en-route to the surface. A number of commercial analog fiber optic links are available, though these products typically have marginal dynamic range and are very expensive. Provided the concept of antenna local triggering and digitization is acceptable, it is much easier to deliver select data to the surface digitally. Within this framework, many commercial data transmission standards are available. Given the ubiquitous nature of the Ethernet standard and the desire to reconstruct and process events in parallel, aided by hardware packet sorting, the Giga-bit Ethernet (GbE) standard is an attractive option. To explore the performance and limitations of this approach a Giga-bit Ethernet Instrument for SalSA Electronics Readout (GEISER) was conceived. A drawing of a single antenna readout is shown in Figure 3. Subsequent sections describe the design, fabrication and test of this GEISER prototype.

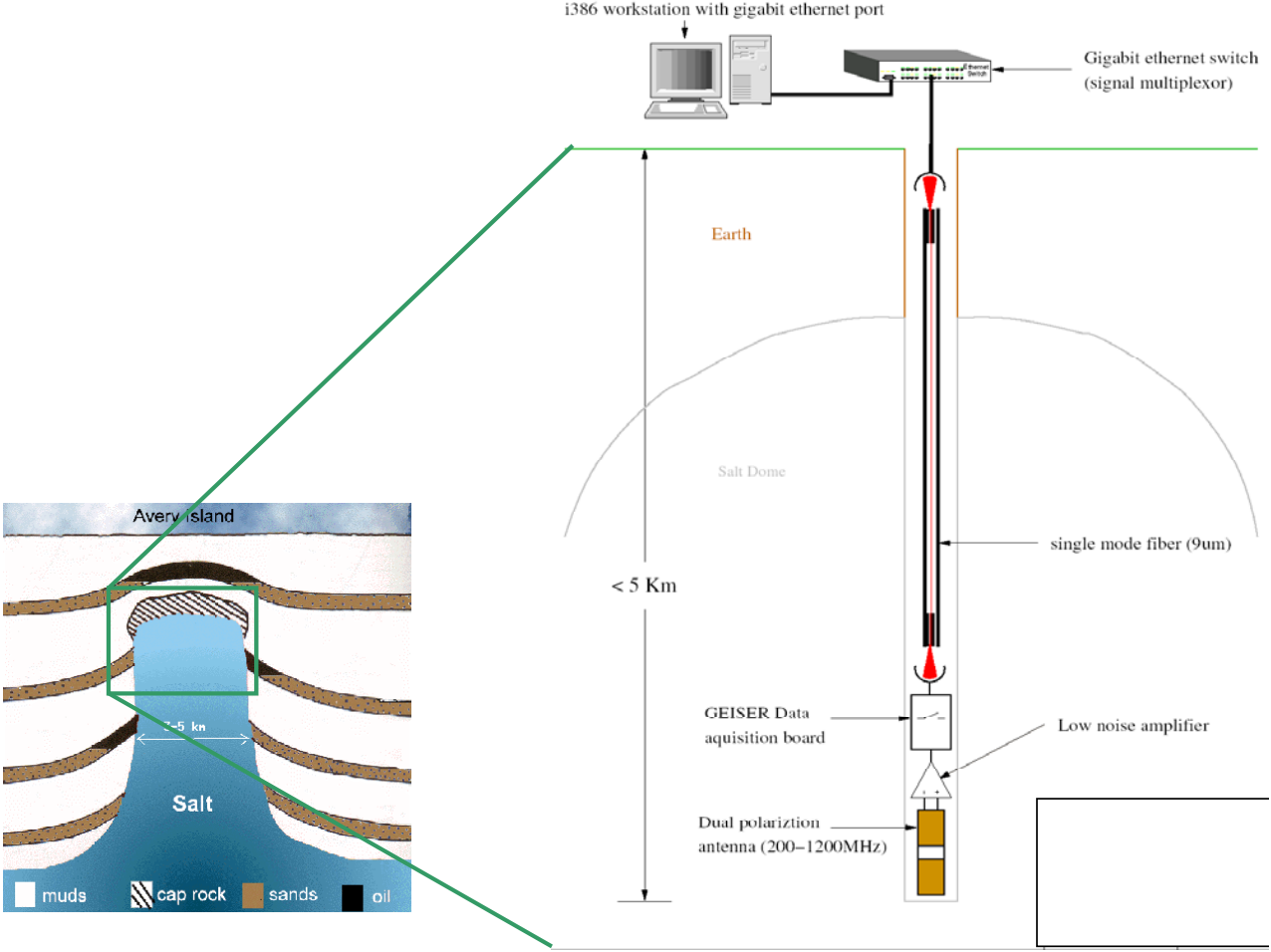

**Figure 3: GEISER readout configuration. As the Askaryan effect impulsive events are completely polarized, the direction of the incident neutrino may be determined by measuring the cross-polarizations. One technique being considered is a slotted pipe dipole, as drawn. Both polarizations are boosted with low-noise amplifiers and processed on the GEISER board. Triggered events are broadcast to the surface where a GbE switch routes packets based on temporal header information to various PCs in a CPU farm for event reconstruction.**

## 2. Specifications

In order to fulfill the functionality dictated by the configuration of Figure 3, the ability to self-trigger is essential. At the time of the design of the initial readout prototype based on Giga-bit Ethernet, only the first fabrication iteration of the STRAW chip (STRAW2) was available. The requirements are summarized below.

Required design features for the GEISER board:

- Better than Nyquist limit sampling of the dual RF antenna inputs at the thermal noise limit over the 100's of MHz of signal BandWidth (BW)
- Ability to trigger on band-limited transients
- Computer controlled RF trigger thresholds
- Continuous monitoring of the RF received power
- Low power, fast triggering
- Precise synchronization with a global timing reference
- Less than 1ms latency

Of these, constraints on simplifying the initial prototype precluded meeting these last two items. A block diagram of the GEISER is shown in Figure 4. A trigger condition is actuated when the received pulse exceeds a DAC selectable

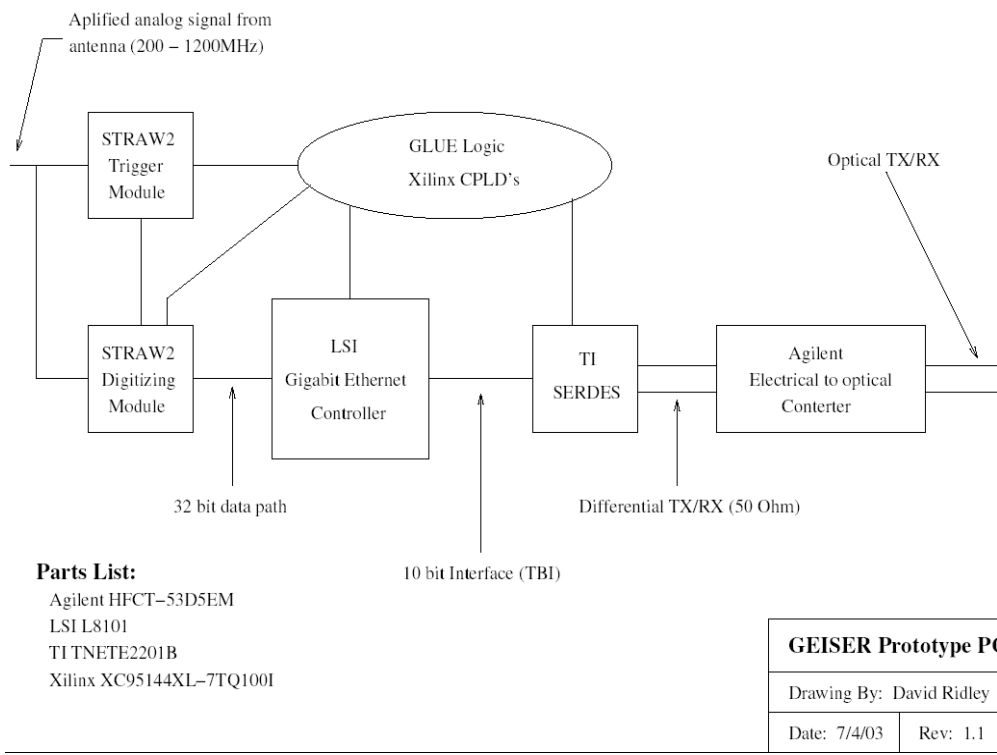

**Figure 4: A block diagram of the GEISER prototype.**

Threshold. Acceptance of a trigger causes analog samples held by the STRAW chip to be digitized. As STRAW2 was used in this prototype, an external ADC was used. The STRAW chip has 16 RF input channels, each of which has 256 switched capacitor array storage cells. In order to maximize the sampling window while maintaining effective sampling rates in excess of the Nyquist limit, each RF input channel is interleave sampled on 4 STRAW inputs. At 2GSa/s effective sampling, this corresponds to a sampling window approximately 500ns wide. As beam test results prove the received RF pulses will be band-limited, the required sampling window is more than adequate to capture the pulse as well as provide pre and post baseline sampling.

As seen in Figure 4, the digitized data is collected from a FIFO located in the STRAW2 Digitizing Module block and formed into frames which are transmitted as blocks by the LSI Controller. The output of this packetizer has been put through 8b-10b encoding for error coding and is transmitted over the 10 bit interface shown to a Serializer/Deserializer (SERDES) which converts the data into a bit serial stream. This bit stream is then put through an Electro-Optical converter which broadcasts the data over a single-mode fiber optic cable. In order to facilitate synchronization of the link, the GbE link is actually bi-directional. This feature will be essential in flow control and timing synchronization amongst antennas in an actual system test and deployment but the downstream link was not explored in this first pass.

## 3. Board Design

As the impetus of this design was to evaluate the functional performance of components and concept, little effort was made to make the board compact. In addition, as functionality that should have been integrated inside the STRAW2 was implemented externally, additional space was required. A photograph of the completed board is provided in Figure 5, with the key components highlighted.

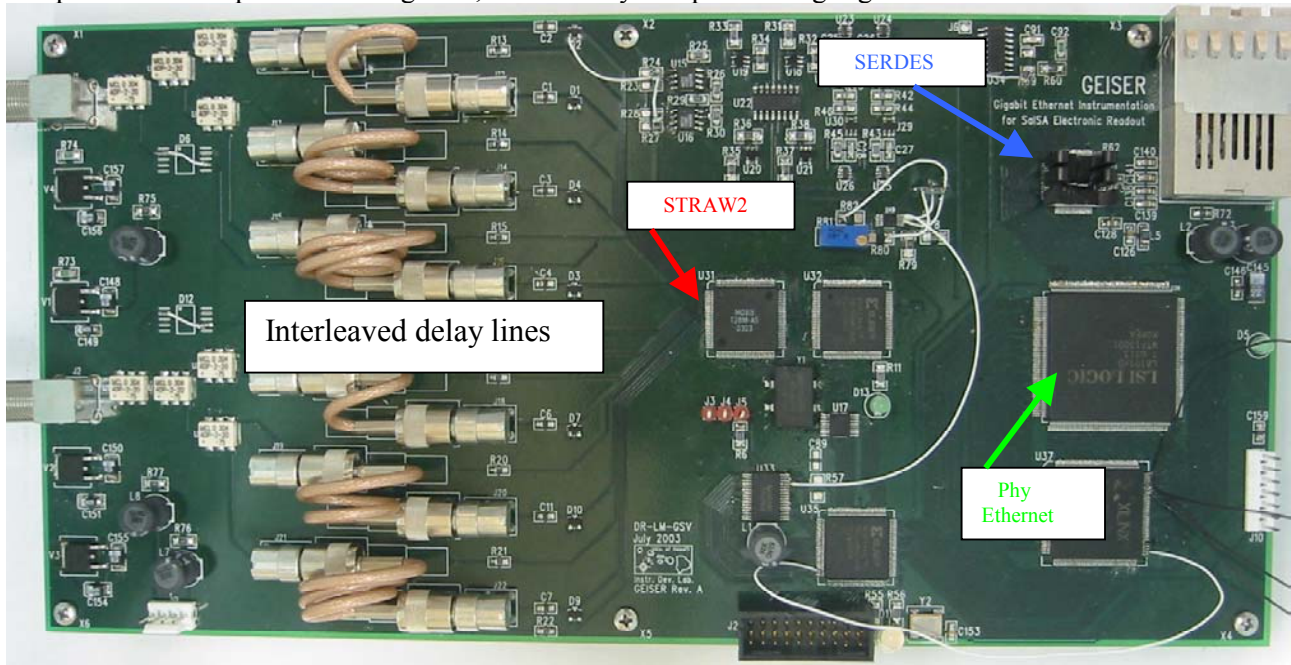

**Figure 5: Photograph of the GEISER prototype with key functional elements indicated. The RF input is supplied on F-connectors with the board a nominal 75 ohm impedance on the inputs. RF splitters are used to fan out the signals to the delay lines as shown. The metal housing at the upper right is the laser Tx-Rx module.**

# 4. Debug and Evaluation

As betrayed by the wires added to the board in Figure 5, some amount of modification to the original design was required in order to fix implementation flaws. Initial testing as part of a student project demonstrated that the board was essentially working. However subsequent, arduous and painstaking debugging was required to fix a serious problem with dropped packets. A valuable lesson learned is that in future revisions, the ability to cache and retransmit dropped packets will be quite valuable. As the external readout of the STRAW2 chip is serial analog, followed by digitization, a significant amount of deadtime is incurred. This was expected, since 16 channels of 256 samples is 4k digitization samples. Given the analog settling time plus the ADC conversion time, about 4ms is required per trigger. With processing overhead, the total latency is about 5ms, corresponding to a maximum trigger rate of about 200Hz. In Table 1 is tabulated the results of a series of measurements performed of the data transfer versus trigger rate.

Table 1: Evaluation of the GEISER performance as a function of trigger rate.

## GEISER Performance Evaluation

| | | | | | | | | |
|---|---|---|---|---|---|---|---|---|
| Trigger rate (number of triggers per second) (R) | 1 | 2 | 5 | 10 | 20 | 50 | 100 | 200 |
| Number of frames received by filter (Nf) | 4.0E+05 | 8.0E+05 | 2.0E+06 | 4.0E+06 | 8.0E+06 | 2.0E+07 | 4.0E+07 | 8.0E+07 |
| Number of frames dropped by kernel (Nd) | 0 | 0 | 0 | 0 | 0 | 3 | 1,479 | 12,479 |
| Number of misordered frames (Nm) | 1 | 1 | 7 | 8 | 29 | 51 | 152 | 2,212 |
| Number of frames with incorrect type (Nt) | 0 | 0 | 0 | 0 | 0 | 0 | 0 | 0 |
| Number of frames with incorrect length (Nl) | 0 | 0 | 0 | 0 | 0 | 0 | 0 | 67 |
| Number of correct packets received (Np) | 44,443 | 88,887 | 222,216 | 444,437 | 888,863 | 2,222,176 | 4,444,291 | 8,886,920 |
| Capturing duration (second) (T) | 4.44E+04 | 4.44E+04 | 4.44E+04 | 4.44E+04 | 4.44E+04 | 4.44E+04 | 4.44E+04 | 4.45E+04 |
| Fraction of frames misordered (Nm/R/400000) | 2.50E-06 | 1.25E-06 | 3.50E-06 | 2.00E-06 | 3.63E-06 | 2.55E-06 | 3.80E-06 | 2.77E-05 |
| Fraction of packets corrupted (1-Np/R/T) | 1.99E-05 | 8.62E-06 | 2.44E-05 | 1.56E-05 | 2.61E-05 | 2.02E-05 | 7.10E-05 | 3.78E-04 |

Input Sine Wave Specification: 25MHz, 700mVpp, 1Cycle per Burst Period

For these measurements a PC running Red Hat Linux Release 9, kernel version 2.4 was used. A custom acquisition program was written by one of the authors (C. Zhu) which is based on the Packet Capture library (PCAP), major version 2, minor version 4. The sampled waveform was a multi-cycle burst 25MHz sine wave with an amplitude matched to full-scale of the STRAW input. As can be seen, many of the failure modes and rates are tabulated. A summary of the overall Packet Error Rate versus trigger rate is shown in Figure 6. A definite trend is seen as the trigger rate approaches complete saturation of the processing capability of GEISER. Note that this is intentionally well below the GbE bandwidth, with the thought of minimizing trigger information transmission latency. At worst the value is still well below 0.1% and in certain circumstances might be acceptable. Addition of packet caching and retransmit capability would greatly improve the robustness of this data pipeline. Careful investigation of possible sources were inconclusive, though this level of error may already be acceptable. For reference, a Packet Error Rate of $5 \times 10^{-4}$ corresponds to an effective Bit Error Rate of about $10^{-9}$. At 200Hz trigger rate, the average sustained transfer rate is approximately 15Mbit/s, with a packet consisting of 9 frames, 8 of which are 1kB and a 9th with 144 bytes.

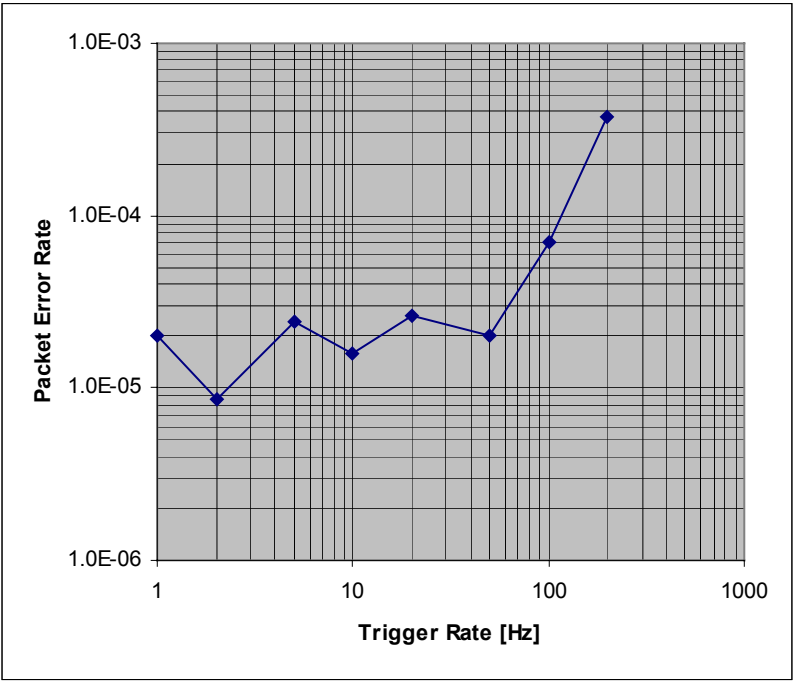

**Figure 6: GEISER transfer statistics summary. Somewhat higher error rates are observed approaching a trigger rate that saturates the processing speed of GEISER readout.**

## 5. Outlook

As the functionality of individual GEISER boards has been demonstrated, the next logical step is to purchase a router and attempt to build events from multiple GEISER boards. Given future resources this will be explored, preferably with a smaller form-factor that will more easily fit down a standard diameter borehole. Addition of packet caching and multiple analog buffering will greatly improve the robustness against data loss. Future versions of the STRAW architecture [6] will have greatly reduced digitization time and there is ample bandwidth yet available on the GbE link to accommodate significantly higher self-trigger rates if needed.